# The Piggy Bank Cryptographic Trope

Subhash Kak

*Abstract:* This paper presents applications of the trope of the locked and sealed piggy-bank into which the secret can be easily inserted but from which it cannot be withdrawn without opening the box. We present a basic two-pass cryptographic scheme that can serve as template for a variety of implementations. Together with the sealed piggy-bank is sent a coded letter that lists and certifies the contents of the box. We show how this idea can help increase the security of cryptographic protocols for classical systems as well as those based on "single-state" systems. More specifically, we propose the use of a hashing digest (instead of the coded letter) to detect loss of key bits to the eavesdropper and use in communication systems where error correction is an important issue.

**Introduction**

The idea of locking a secret in a box and letting it be carried to the destination by an unreliable courier (Figure 1) (where it is unlocked by the recipient who has the key to unlock the box) is at the basis of most cryptographic protocols. This scheme assumes that the key has somehow been transported to the recipient in advance of the communication. The lock of the box is protected by placing a seal across it that ensures that it is not tampered with by the courier.

In the case of the use of this scheme in data communication, the key may be transmitted over a side channel. If the rate at which the key is transmitted over the side channel equals the data rate, then this constitutes the unbreakable one-time pad [1].

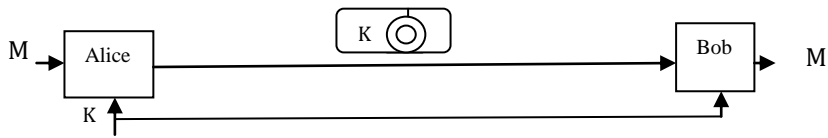

Figure 1. Sending a secret in a locked box
(unlocking needs key K)

Another idea is that of the three-stage protocol (Figure 2) which can be used when the recipient does not possess a copy of the key. This requires that both parties use locks and it is assumed that the locks are protected by tamper-proof seals of the two parties. In this protocol Alice puts the secret in a locked box which is transported to Bob who puts his own lock on the box and sends it back to Alice who unlocks her lock and resends the box to Bob who then unlocks his lock. This protocol ensures that both Alice and Bob can check that their locks have not been tampered with. Amongst other applications, this idea is at the basis of the three-stage quantum cryptography protocol [2].

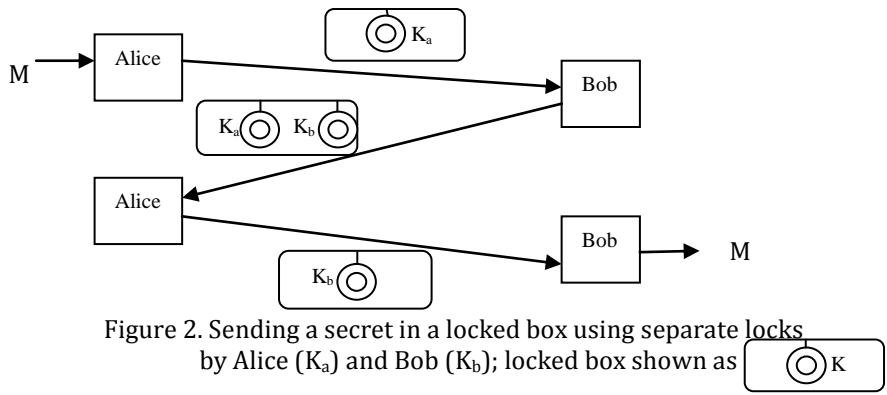

Figure 2. Sending a secret in a locked box using separate locks by Alice ($K_a$) and Bob ($K_b$); locked box shown as

Although the locked box is the most popular foundational unit of traditional secure systems, it is not the only one. Another basic unit, with lesser popularity in formal arrangements but equally great popularity in informal systems, is that of the piggy-bank (Figure 3) in which coins or money can be easily inserted but not withdrawn without access to the key with which it has been locked.

*Jehoiada the priest took a chest, and bored a hole in the lid of it, and set it beside the altar, on the right side as one cometh into the house of the LORD: and the priests that kept the door put therein all the money that was brought into the house of the LORD. --*
2 Kings 12.9

We propose that use of such a locked box with a receptacle for insertion of money (or secrets) can be the model for cryptographic systems. Although used for collecting money at a public location, the box was sometimes moved to another location for counting the money and bills.

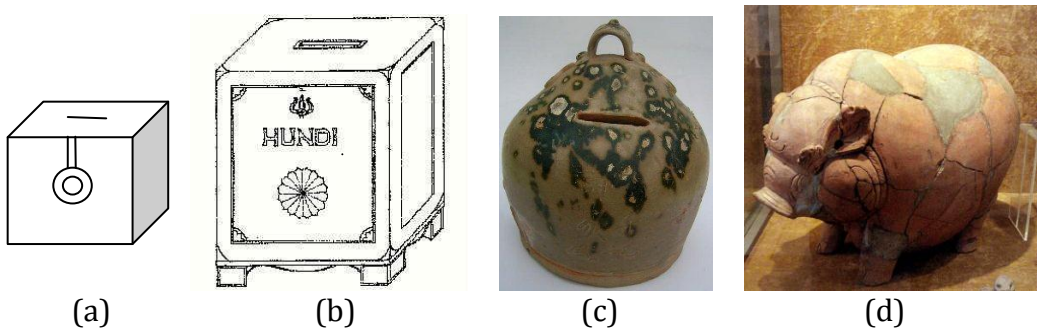

(a)　　　　　　(b)　　　　　　(c)　　　　　　(d)

Figure 3. (a) A piggy bank; (b)A temple money box; (b) Tang Dynasty piggy bank from China; (c) piggy bank from Majapahit Indonesia (14th-15th century) (National Museum of Indonesia, Jakarta)

**The piggy bank trope**
Bob sends an empty locked piggy bank to Alice. When she receives it, Alice deposits the secret (money, bills, and jewels) into the box together with the decryption key of a coded letter. In addition, she prepares a letter to be sent separately. The piggy bank and the letter are sent back to Bob.

The letter is required to authenticate the contents of the locked piggy bank box. It cannot be in plaintext because the content list itself is a secret. The letter is needed to establish the identity of the person who has sent the secret (that is Alice) and this may carry an additional secret.

Bob opens the box, obtains the secret, and also reads the coded letter which has further details of the secrets in it.

This protocol is described in Figure 4.

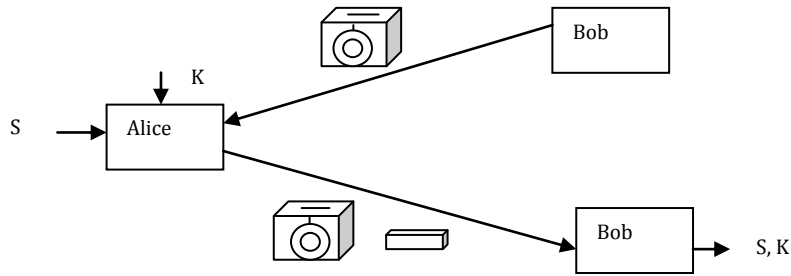

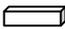

Figure 4. The piggy bank cryptographic trope; the secret letter is represented by 

The idea of sending two pieces of secret information in partitions was used earlier by the author in the different context of visual cryptography [3]. In the case where Bob's ability to read the "secret letter" is limited, he will be able to obtain only one of the two secrets.

The piggy bank trope can be implemented in many variations by making further assumptions about the system. Here we provide a few where standard primitives are employed.

**Protocol 1**
In this implementation of the piggy bank protocol for data, Bob obtains both the secrets K and S. The protocol consists of three steps of Figure 5:

*Step 1.* Bob starts with a random number R and the piggy bank transformation is represented by a one-way transformation $f(R) = R^e \mod n$, where n is a composite number with factors known only to him; e is the publicly known encryption exponent.

*Step 2.* Bob sends f(R) to Alice who multiplies it with her first secret S. Alice sends $S(R)^e + K \mod n$ to Bob in one communication and $f(S) = S^e \mod n$ in another communication.

*Step 3.* Bob uses his secret inverse transformation to first recover S and having found it he can recover K.

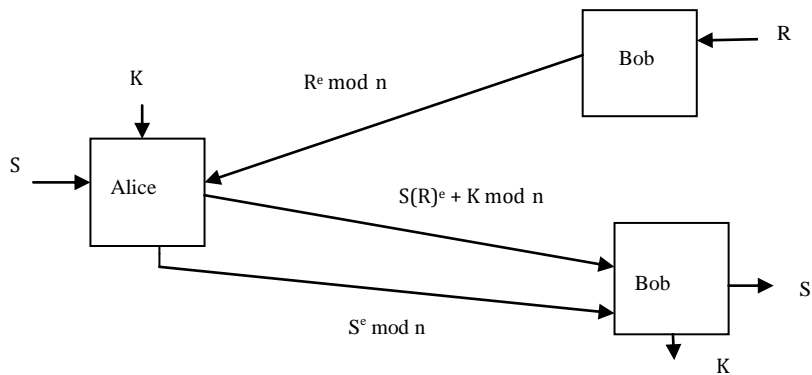

Figure 5. The piggy bank protocol 1 for communications

EXAMPLE 1. Let n= 51 and the public encrypting exponent is e=3 (with the secret decrypting algorithm being 11 since $3 \times 11 = 1 \mod \varphi(51)$ ). Bob chooses random R=13 and sends $13^3 \mod 51 = 4$ to Alice.

Alice's random secrets are S=5 and K=29. Alice computes $4 \times 5 + 29 = 49$ and sends it and also $5^3 \mod 51 = 23$ to Bob.

Bob uses his secret decryption exponent to recover S: $23^{11} \mod 51 = 5$. Thus $5 \times 4 + K = 49$, from which he recovers K.

**Three Variations on Protocol 1:**

These variations require correspondingly appropriate actions by Bob.

1. Take R=1. This means that Alice sends $S+K \mod n$ and $S^e \mod n$.
2. Alice sends $R^{eS} \mod n$ and $S^e \mod n$ to Bob.
3. Bob sends R to Alice who, in turn, sends $S^e R+K \mod n$ and $S^e \mod n$ to Bob.
4. Bob sends R to Alice who, in turn, sends $SR+K \mod n$ and $K^e \mod n$ to Bob.

**Protocol 2**
This implementation where Bob obtains only one of the two secrets. The two parties also obtain an additional shared random number. The protocol consists of the following steps:

*Step 1.* Bob starts with a random number R and the piggy bank transformation is represented by a one-way transformation f. The transformation could be exponentiation of a publicly announced generator g of the elements of the multiplicative group modulo p, which is a prime.

*Step 2.* Bob sends f(R) to Alice who conjoins it with the secret S that she wishes to send to Bob and then performs the transformation f. It is assumed that f(S*f(R)) = f(SR), so as to see the operations performed by Bob and Alice to be similar. Now Alice sends f(S*f(R))+K to Bob which is equivalent to f(SR)+K as well as f(S) separately.

*Step 3.* Bob performs f(R*f(S)) which is equivalent to f(SR) since he knows the value of R. Now he subtracts it from f(SR)+K and, thereby, obtains K.

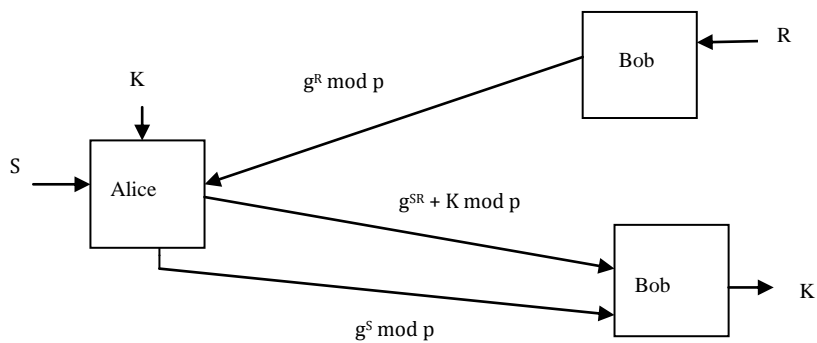

Figure 6. The piggy bank protocol 2 for communications; f(Sf(R))=f(SR)

In a variant of this protocol, Alice sends K×f(Sf(R))+K and f(S) to Bob, although this does not provide any special advantage.

An implementation of the protocol is given in Figure 6.

EXAMPLE 2. Let $f(x)=g^x$ mod p, where p is prime. Let p=37, g=2, and R=11. Therefore, Bob sends $2^{11}$ mod 37 = 13.
    Let Alice's secrets be S=3 and K=10. Alice generates $13^3$ mod 37 = 14 and sends 14 + K that equals 24 and also $2^S$ mod 37= 8.
    Bob computes $(2^S)^{11}$ mod 37 and obtains 14. Subtracting this from 24, he obtains the secret K to be equal to 10.
    Alice and Bob also come to share the random number $g^{SR}$ mod p that could be used for some other cryptographic purpose.

Like protocol 1, protocol 2 can be implemented in other variations including one of the secrets is conjoined with the other encrypted terms in a multiplicative way (rather than the additive way shown here). It may also be generalized. The secret letter may be replaced by a hashing digest in certain situations (as in the application to quantum cryptography described below).

**Hashing digests in quantum cryptography**

The piggy bank trope may be applied to quantum cryptography [4]-[6] although this cannot be done in quite the same way as in the protocols above. Specifically, we can use the second "letter" communication from Alice to Bob to send a hashing digest of the key to determine if the bits have been correctly received.

Consider the communications [7]-[9] and signal-to-noise (SNR) perspectives on cryptography [10]. The BB84 protocol requires that single photons be sent by Alice to Bob [11], but the ability to receive single photons means that the SNR ratio for the receiver is infinite and the channel is fully protected. With such transmission requirements, there may very well be no need to use encryption!

In BB84 we could do away with complicated error correction, like "cascade", to counter the eavesdropper if the resources for computing commonly available on the network are harnessed to send side information.

In cascade, the sifted key bits are divided into blocks and then both parties announce the parity of each of these blocks on a public channel. If Alice's parity for a block differs from Bob's, it is clear that there are an odd number of errors in that block. The search for these errors is done recursively, by dividing the block into smaller ones, until only an even number is contained in that block. When the blocks have been processed, the bits are shuffled and the procedure repeated. This is done a number of times, so that the probability that the remaining key contains an error is very low.

Instead of the cascade procedure, a cryptographically strong hash digest of the raw key can be sent to Bob to ascertain if the eavesdropper has siphoned off any photons or if noise has led to any errors. This digest may be sent separately to the destination quite like the "coded letter" of Figure 4. If the digest generated by Bob doesn't match the one he has received from Alice, he asks for a retransmission of the bits.

In the use of hashing as a resource, the hash digest may be shared amongst the users on a side-channel since it is assumed that bandwidth is not limited. The BB84 protocol assumes that the data is being transmitted by single objects (photons) for if more than one photon is transmitted for each bit, the eavesdropper can siphon off the superfluous bit to obtain partial information about the key being transmitted. Of course, not all quantum cryptography systems use single photons as evidenced by the three-stage protocol using random rotations [2]. But even here the number of photons in each communication must be restricted so that the eavesdropper does not have information to determine the polarizations in each of the three links.

It is true that the use of the hashing digest will not prevent the eavesdropper from disrupting the communication.

**Discussion**

This paper has shown how the trope of the piggy bank can have cryptographic applications in communications and key-distribution systems. We have provided examples of basic use in classical and quantum cryptography. Further variations on the protocols provided in this paper may be easily developed.